\documentclass[manuscript]{aastex}
\usepackage{amsmath}

\slugcomment{\textbf{Astrophysical Journal Letters, in press}}

\shorttitle{Active Asteroid P/2020 O1}
\shortauthors{Kim et al.}

\begin{document}

\title{Hubble Space Telescope Observations of Active Asteroid\\
 P/2020 O1 (Lemmon-PANSTARRS)}

\author{Yoonyoung Kim$^1$,
David Jewitt$^{2}$,
Jessica Agarwal$^1$,
Max Mutchler$^3$,
Jing Li$^2$ and
Harold Weaver$^4$  
}

\affil{$^1$ Institute for Geophysics and Extraterrestrial Physics, TU Braunschweig, 38106 Braunschweig, Germany\\
$^2$ Department of Earth, Planetary and Space Sciences,
UCLA, Los Angeles, CA 90095-1567\\
$^3$ Space Telescope Science Institute, Baltimore, MD 21218 \\
$^4$ The Johns Hopkins University Applied Physics Laboratory,  Laurel, Maryland 20723
} 

\email{yoonyoung.kim@tu-bs.de}

\begin{abstract}

We present Hubble Space Telescope observations of active asteroid P/2020 O1 taken to examine its development for a year after perihelion.
We find that the mass loss peaks  $\lesssim$ 1 kg s$^{-1}$ in 2020 August and then declines to nearly zero over four months.
Long-duration mass loss ($\sim$180 days) is consistent with a sublimation origin, indicating that this object is likely an ice-bearing main-belt comet.
Equilibrium sublimation of water ice from an area as small as 1580~m$^2$ can supply the observed mass loss.
Time-series photometry shows tentative evidence for extremely rapid rotation 
(double-peaked period $<$ 2 hr) 
of the small nucleus (effective radius $\sim$420~m).
Ejection velocities of 0.1~mm particles are comparable to the 0.3 m s$^{-1}$ gravitational escape speed from the nucleus, while larger particles are ejected at speeds less than the escape velocity.
These properties are consistent with the sublimation of near-surface ice aided by centripetal acceleration. 
If water ice sublimation is confirmed, P/2020 O1 would be the icy asteroid with the smallest semimajor axis (highest temperature), setting new bounds on the distribution of ice in the asteroid belt.

\end{abstract}


\section{INTRODUCTION}

Active Asteroid P/2020 O1 (hereafter ``O1'') was discovered on UT 2020 July 20 (Weryk et al. 2020), shortly after passing perihelion at 2.329 AU on UT 2020 May 03.   With semi-major axis $a$ = 2.647 AU, eccentricity $e$ = 0.120 and inclination $i$ = 5.2$\degr$, it lies in the middle of the asteroid belt, closer to the Sun than the so-called ``main-belt comets'' (MBCs), the subset of the active asteroids near $a$ = 3 AU that are thought to be ice sublimators (Hsieh \& Jewitt 2006; Jewitt \& Hsieh 2022).  The Tisserand parameter with respect to Jupiter, $T_J$ = 3.38, is strongly asteroid-like and its orbital elements fall outside the range into which capture from elsewhere might have occurred (Hsieh \& Haghighipour 2016; see their Figure~6). 

In this paper we report observations of O1 from the Hubble Space Telescope taken to sample its morphological and photometric development at the highest angular resolution.
We compare the observations with a sophisticated Monte Carlo model of  dust dynamics to determine the cause of the activity for an object located in the warm middle-belt.

\bigskip \bigskip
\section{OBSERVATIONS}

Observations with the HST were taken under target-of-opportunity program GO 16308 (two orbits) and continued with mid-cycle time under GO 16463 (five orbits).
We used the UVIS channel of the WFC3 camera with the broadband F350LP filter (effective wavelength $\sim$5846\AA, full width at half maximum (FWHM) $\sim$4758\AA) in order to provide maximum sensitivity to faint coma. The pixel scale and the field of view are 0.04\arcsec~pixel$^{-1}$ and 80\arcsec$\times$80\arcsec, respectively, where the asteroid is centrally located.
A journal of observations is shown in Table \ref{geometry}.

In each HST orbit, we obtained six exposures of 230--260~s duration (1380--1560~s per orbit).
The individual HST images are strongly affected by cosmic rays. A clean image was obtained by computing the median of the six images from each orbit.
For the August 24 data, only two out of six exposures were useful due to guide star issues.
We removed cosmic rays in the August 24 composite image by hand, replacing them with the nearest good pixels.
The last observations, on UT 2021 July 22, were obtained from three closely-spaced HST orbits, while observations on the remaining dates were each obtained from a single HST orbit.
Special observations were targeted on UT 2020 September 17 as the Earth passed through the projected orbital plane of O1 (the out-of-plane angle was -0.03\degr).  This viewing geometry provides a strong constraint on the out-of-plane distribution of dust and hence the perpendicular ejection velocity.

\section{RESULTS}

\subsection{Morphology}
\label{morphology}

\begin{figure}
\epsscale{1.0}
\plotone{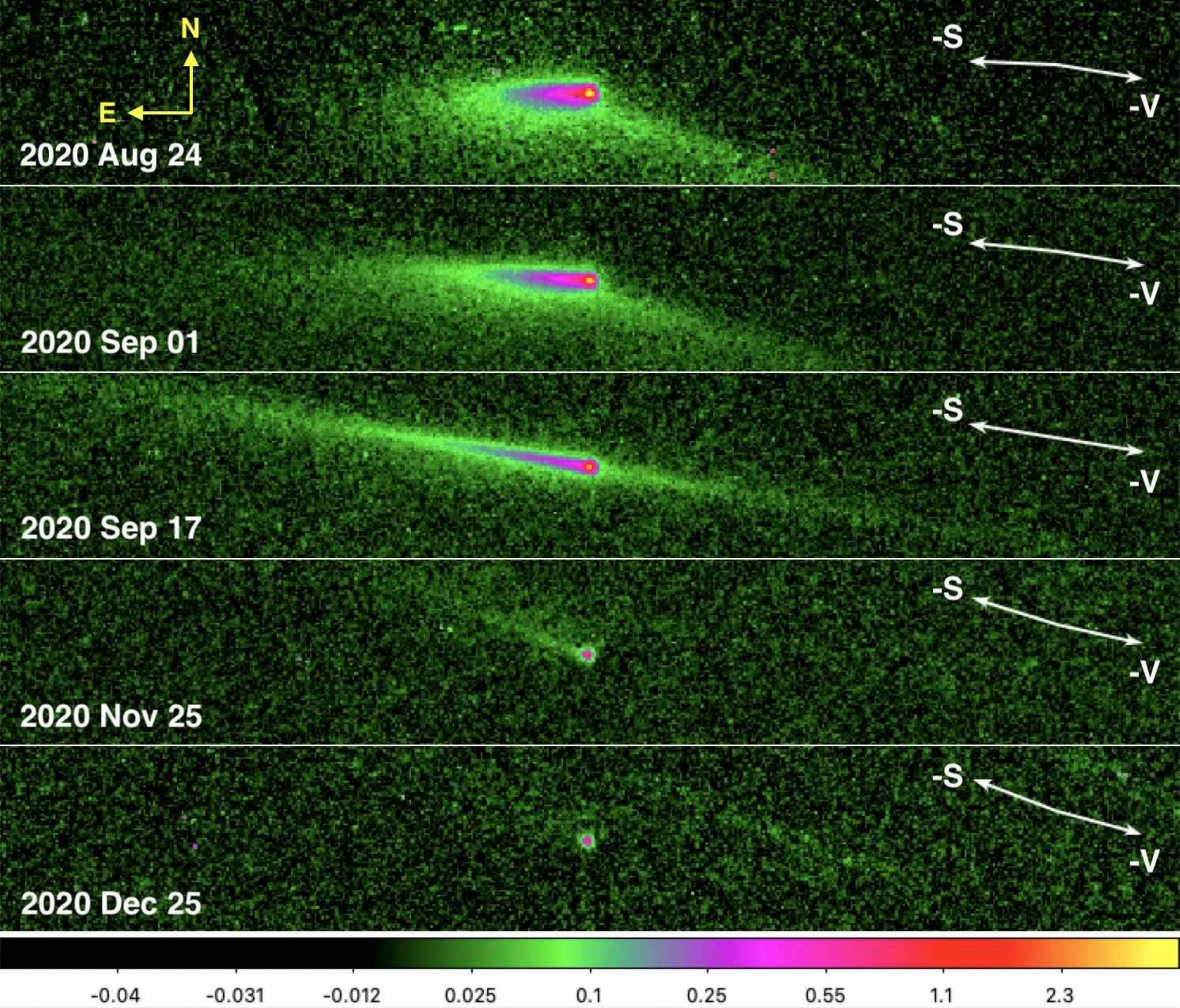}
\caption{Composite HST images of active asteroid P/2020 O1 marked with UT dates of observation.
A color bar (logarithmic scale), the projected anti-solar direction ($-S$) and the negative heliocentric velocity vector ($-V$) are indicated.
Each panel shows a region $18.0\arcsec \times 2.5\arcsec$ in the standard orientation: celestial north points up and east points to the left.
 \label{images}}
\end{figure}

Composite images of O1 for each date of observation are shown in Figure \ref{images}, with direction vectors showing the anti-solar and negative heliocentric velocity directions.
The morphology of O1 changes systematically over time, from showing a fan-shaped tail in 2020 August to a thin, linear tail as the Earth passes through the orbital plane in September, toward a point-like appearance in 2020 December and 2021 July.
Observations   show the dust tail, consisting of recently ejected, relatively small, radiation pressure-swept particles to the east and larger, older particles along the projected orbit to the west. 
The position angle of the east tail changes clockwise following the changing anti-solar direction.
Images from 2020 August and September show a unique morphology where the axis of the tail to the west does not intersect the nucleus.

\begin{figure}
\epsscale{0.95}
\plotone{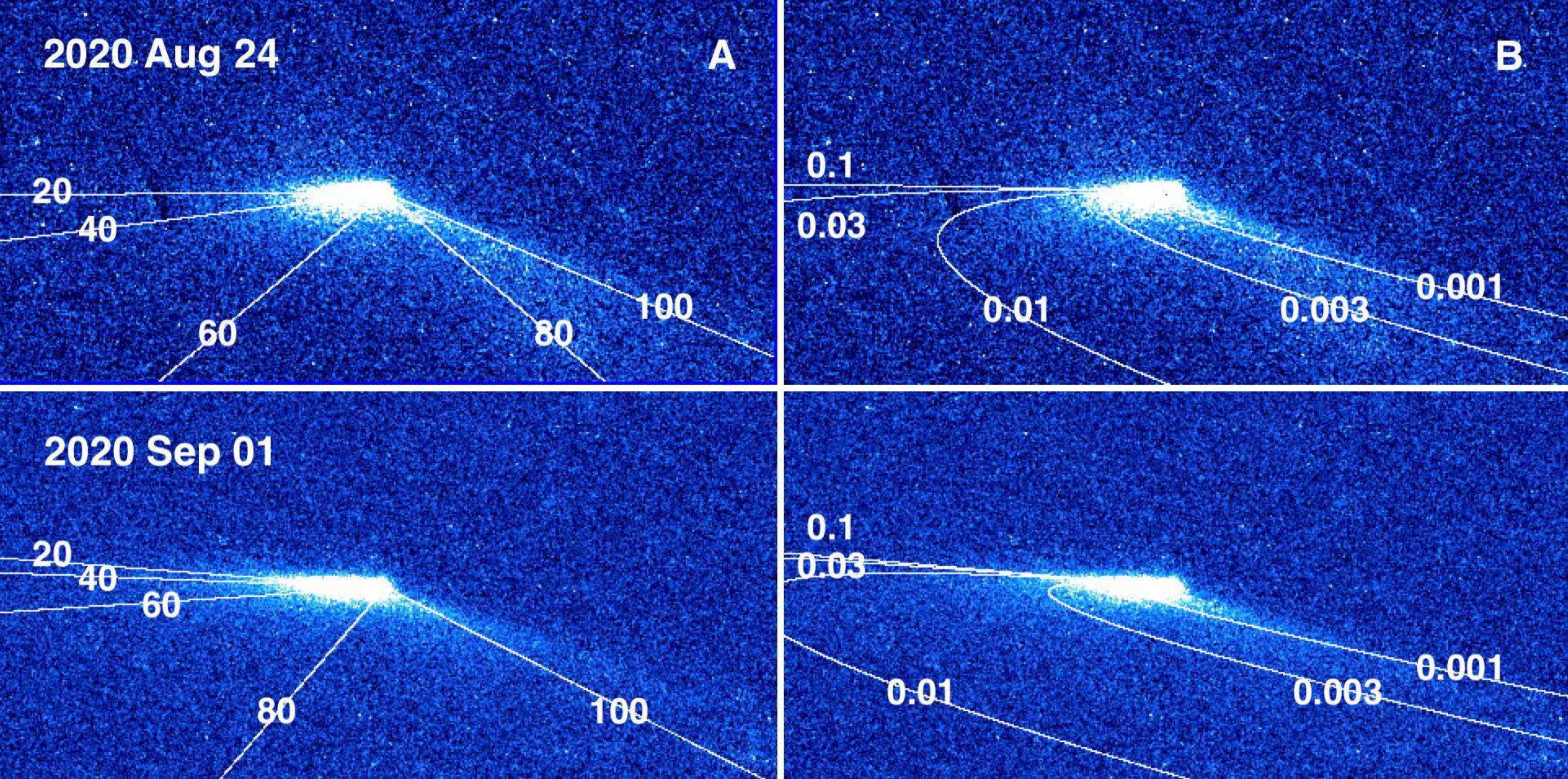}
\caption{(A) synchrones computed for ejection 20, 40, 60, 80, and 100 days prior to the date of observation and (B) syndynes, showing the paths of particles with $\beta$ = 0.1, 0.03, 0.01, 0.003, and 0.001.
Celestial north points up and east points to the left.
\label{synsyns}}
\end{figure}

The motion of a dust particle of radius, $a$, is controlled by $\beta$, the ratio of radiation pressure acceleration to solar gravity.
For a spherical dielectric particle, $\beta$ is approximately given by $\beta = 0.57/\rho_d a_{\mu m}$ (Bohren \& Huffman 1983), where $a_{\mu m}$ is the particle radius in microns and $\rho_d$ = 1 g cm$^{-3}$ is the assumed particle density.
We assumed that ejected dust particles are compact in shape and optically large.
For each epoch of observation, we computed syndyne/synchrone (Finson \& Probstein 1968) trajectories.
Figure \ref{synsyns}(a)  shows syndynes, the loci of particles of a given $\beta$ ejected with zero velocity at different times.  Figure \ref{synsyns}(b) shows synchrones, which are the loci of particles having different $\beta$ but which are ejected at a given time.
The Figure shows that the direction and slight curvature of the tail are best matched by syndynes with $\beta \sim$ 0.0005 to 0.007.
The easternmost extent of the detected tail corresponds to particles having $\beta_\mathrm{max} = 0.007$ (radius $\sim$0.08 mm).
The ``curved'' northern edge of the western tail corresponds to particles having $\beta_\mathrm{min} = 0.0005$ (radius $\sim$1.14 mm), which therefore constrains the maximum particle radius.
We take the geometric mean, $a \sim$ 0.3 mm, as the nominal grain size.  
The western edge of the detected tail corresponds to particles ejected in early 2020 May (synchrones $\gtrsim$100 days prior to the date of observation), indicating the onset of activity near perihelion.

\subsection{Photometry}
\label{photometry}

We obtained photometry from each composite image (Figure \ref{images}) using a set of five circular apertures having fixed radii from 500 to 8,000~km, when projected to the distance of O1.
The sky background was determined within a concentric annulus with inner and outer radii of 20\arcsec~and 25\arcsec, respectively.
Flux calibration was performed using the online WFC3 Exposure Time calculations for a solar-type source in the F350LP filter.
We corrected the apparent magnitudes, $V$, to absolute magnitudes, $H$, using

\begin{equation}
H = V - 5\log_{10}(r_H \Delta) + 2.5\log \Phi(\alpha).
\label{absolute}
\end{equation}

\noindent in which $r_H$ and $\Delta$ are the heliocentric and geocentric distances, respectively. $0 \leq \Phi(\alpha) \leq 1$ is the phase function at solar phase angle $\alpha$.
In the absence of an empirical determination, we used the phase function formalism of Bowell et al.~(1989) with parameter $g$ = 0.15, as appropriate for a C-type object and $g$ = 0.25, for an S-type object.  At the largest phase angles of our observations ($\alpha$ = 22.6\degr, Table \ref{geometry}), the difference between assumed C-type and S-type phase corrections is $\sim$0.1 mag.

We additionally identified and analyzed archival data from the ZTF (Zwicky Transient Facility; Bellm et al. 2019) obtained on UT 2020 August 12 using ZTF $g$ and $r$ band filters.
ZTF photometry  indicates color $g-r$ = 0.48$\pm$0.10 on UT 2020 August 12 within a circular photometry aperture of projected radius 5.0\arcsec.  This is a one-time color measurement and there is uncertainty, however, we assume that O1 emitted dust with optical color more C-like than S-like.  In the rest of the paper, we assume a C-like phase function, albedo, and density.

The absolute magnitudes are converted into the effective scattering cross sections, $C_e$ (km$^2$), by
 
\begin{equation}
C_e = \frac{2.24\times10^{16} \pi}{p_{V}} ~10^{0.4[m_{\odot, V} - H_{V}]}
\label{area}
\end{equation}

\noindent where $p_V$ is the geometric albedo and $m_{\odot, V}$ = -26.77 is the apparent V magnitude
of the Sun.  We assume $p_V$ = 0.05, consistent with the albedos of MBC nuclei (Hsieh et al. 2009).
For each date and aperture radius, $V$, $H$, and $C_e$ are given in Table \ref{phot} with their photometric uncertainties.  The actual uncertainty on $C_e$ is larger because of the assumed phase correction and geometric albedo.

\begin{figure}
\epsscale{1.2}
\plotone{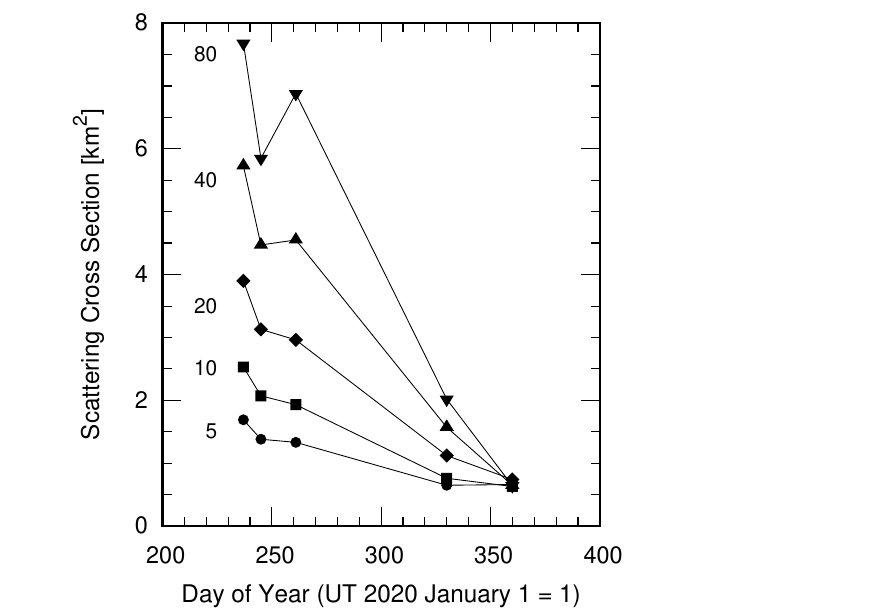}
\caption{Scattering cross-section as a function of time, expressed as Day of Year (DOY=1 on UT 2020 January 1).
The radii of the apertures (in units of 10$^2$ km) are indicated.   \label{ce}} 
\end{figure}

In Figure \ref{ce}, the scattering cross-section is seen to decrease until the  observation on UT 2020 December 25 (DOY 360) as O1 moves away from perihelion.  The bump within the 8,000~km radius aperture on UT 2020 September 17 (DOY 261) is affected by imperfect removal of scattered light and trailed field objects.   The central aperture flattens after UT 2020 November 25 (DOY 330) and remains relatively constant, indicating that dust production has ceased and that dust has left the immediate environment of the nucleus due to solar radiation pressure.
The December 25 image appears point-like, and the scattering cross-section at larger apertures appears consistent with the central aperture.
This also indicates that dust production has completely ceased at the time of observation.

About 8 months after the inferred  termination, observations on UT 2021 July 22 ($r_H$\,=\,2.745 AU) reveal the object in an inactive state and a mean absolute magnitude $H_V$\,=\,19.25$\pm$0.13.  Substituting into Equation (\ref{area}) we obtain the nucleus cross-section $C_n$\,=\,0.55$\pm$0.07 km$^2$ and the equivalent circular radius  $r_n$ = ($C_n/\pi)^{1/2}$ = 0.42$\pm$0.03 km.  The nucleus cross-section is $\sim$0.1 km$^2$ smaller than the cross-section obtained for the same sized aperture on November 25 and December 25, indicating that the large, slow particles remaining near the nucleus moved slowly over $\sim$8 months.

\subsection{Nucleus Photometry and Rotation}

The central aperture is most  sensitive to the brightness of the nucleus.
Although photometry reveals the presence of particles that have not been completely removed, we first investigated photometric variations in the nucleus using November 25 and December 25 images.
Photometry of the nucleus was obtained using projected circular apertures 0.2\arcsec~in radius, with sky background  determined within a surrounding annulus having inner and outer radii 0.2\arcsec~and 0.4\arcsec, respectively (Table \ref{nucleus_phot}).  The measurements show variations in the brightness that appear to be non-random, with a peak-to-peak amplitude  $\Delta m$ = 0.28~mag.     If attributed to the rotational variation of the cross-section of an $a \times b$ ellipsoid, then $a/b$ = 10$^{0.4 \Delta m}$ = 1.3, with $a \times b$ = (0.48 $\times$ 0.37) km. 

\begin{figure}
\epsscale{1.2}
\plotone{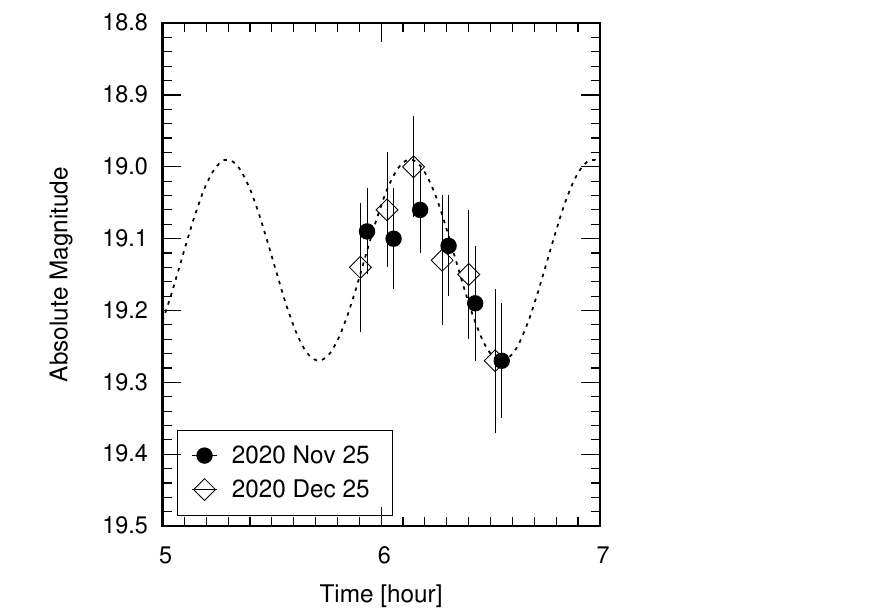}
\caption{Lightcurve of O1 in data from UT 2020 November 25 and UT 2020 December 25. 
The December data have been shifted assuming a double-peaked period of 1.67 hr.
A sine curve with a period of 0.83 hr was plotted with the dashed line.
\label{lightcurve}}
\end{figure}

A phase dispersion minimization estimate of the period in the O1 data gives a best-fit single-peaked period $P_0$ = 0.83 hr,
while a secondary peak in the periodogram indicates another possible period of $P_0$ = 0.95 hr.
Assuming that the lightcurve results from a rotating ellipsoidal body, we obtained a double-peaked period of $2P_0$ = 1.67 hr, or alternatively $2P_0$ = 1.90 hr.
The data suggest the possibility that O1 is rotating close to rotational instability ($<$ 2 hr).
In addition, three closely-spaced HST orbits on UT 2021 July 22 were secured specifically to provide a timebase sufficient to assess short-term variations in the scattered light.  However, the photometric uncertainty of the apparent magnitude $V \lesssim 25$ point source at a 230~s exposure was about $\gtrsim$0.15 mag, comparable to the amplitude of the lightcurve. We judge that the photometry from this date did not reach a quality sufficient to accurately measure the rotational lightcurve.
Pending the acquisition of better photometry needed to confirm the periodicity in Figure \ref{lightcurve}, we leave open the role of rotation in affecting the activity of O1.

\subsection{Dust Profiles}
\label{profile}

\begin{figure}
\epsscale{0.95}
\plotone{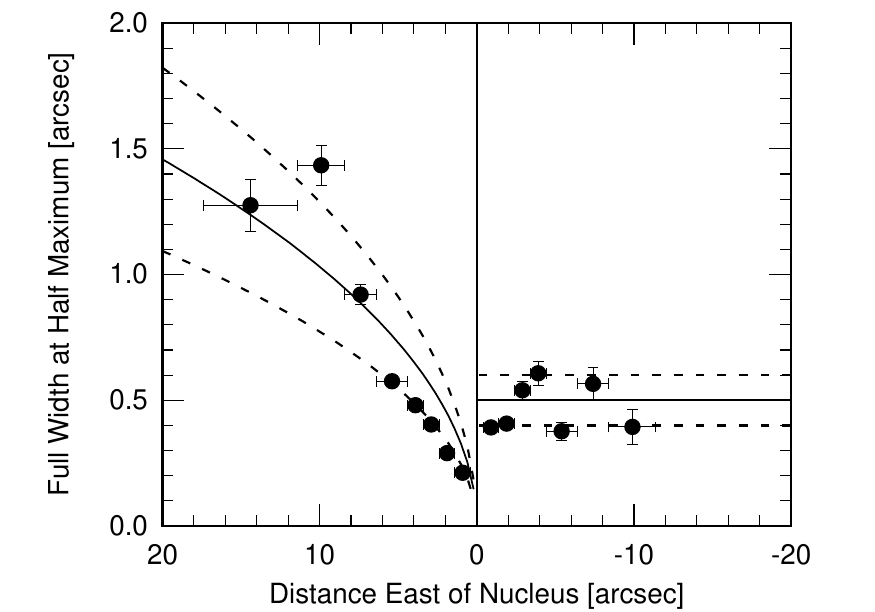}
\caption{FWHM of the dust tail as a function of the angular distance from the nucleus, observed at plane-crossing on UT 2020 September 17.
Best-fit lines (Equation \ref{width}) to the east of the nucleus indicate ejection velocities 4$\pm$1 m s$^{-1}$ (for $\beta=1$ particles).
 \label{fwhm}}
\end{figure}

Observations on UT 2020 September 17 were taken as the Earth passed through the projected orbit plane of O1, and offer a powerful constraint on the out-of-plane distribution of dust.  
We used the September 17 composite image to determine a series of surface profiles cut perpendicular to the tail.
Figure \ref{fwhm} shows the FWHM measurements, $w_T$, as a function of the projected angular distance from the nucleus, $\theta$.
Vertical error bars in the figure show uncertainties in the FWHM measurement, while horizontal bars indicate the width of the segment used to make the profiles.
Our data show a very narrow west tail (FWHM $\sim$ 0.5\arcsec) and a thicker east tail that gradually widens as the distance from the nucleus increases.

The width of the tail, $w_T$, is related to the distance from the nucleus, $\ell_T$, by

\begin{equation}
V_{\perp} = \left(\frac{\beta g_{\odot}}{8 \ell_T} \right)^{1/2} w_T
\label{width}
\end{equation}

\noindent  where $V_{\perp}$ is the perpendicular ejection velocity and $g_{\odot}$ is the local solar gravitational acceleration (Jewitt et al.~2014).  
For simplicity, we assume $\ell_T \propto \theta$ and neglect projection effects. 
This assumption is less accurate for older particles to the west of the nucleus, as can be seen from the curved syndynes in Figure \ref{synsyns}.
The syndyne trajectories first head east and then turn west, and do not simply follow $\ell_T \propto \theta$ to the west of the nucleus.
Thus, we measure the ejection velocity for the younger, eastern tail only.
We show Equation (\ref{width}) fitted to the east tail  in Figure \ref{fwhm}, finding $V_{\perp}$ = 4.0$\pm$1.0 m s$^{-1}$ for $\beta=1$ particles.
Within the uncertainties, we take $V_{\perp} \sim$ 4.0 $a_{\mu m}^{-1/2}$ m s$^{-1}$  as the dust ejection velocity forming the east tail.

\section{DISCUSSION}

\subsection{Dust Dynamical Model}
\label{model_text}

While the presence of a flared distribution on the east side of the nucleus is expected, the approximately constant FWHM of the dust to the west of the nucleus is more surprising (Figure \ref{fwhm}).
Comparison with the syndyne/synchrone model suggests that the west tail consists of either large (slow) particles ($\beta<0.001$) or old particles emitted before UT 2020 May 30, or both.  

As suggested by the syndyne analysis (Figure \ref{synsyns}), the largest particles should have $\beta_\mathrm{min} \sim 0.0005$ to form a curved west tail.
Several possibilities exist for the asymmetry of east-west tail width (dust velocities):

\begin{enumerate}

\item Time variable active fraction. Assuming dust particles are ejected in a sunward cone with a half-angle $\omega$, the tail width is controlled by $V_{\perp} = V_0 \sin(\omega)$. An active fraction increasing with time may explain the observed tail width.

\item Time variable dust ejection velocity. If the dust ejection velocity was initially small and then gradually increased, it could explain the observed tail width.

\item Decelerated terminal dust speeds in the presence of nucleus gravity or rapid rotation.

\end{enumerate}

Considering the very small active fraction at the activity peak (Section \ref{mechanism}), explanation (1) is less plausible. Explanation (2) is also unnatural as it requires a significantly smaller dust ejection velocity near  perihelion (opposite to our expectations, $V \propto r_H^{-1}$). 
In the classical comet model (Whipple 1951), the solution to the velocity-radius relation involves nuclear gravity, setting a critical size that particles are too heavy to be ejected by gas drag.
In this picture, Explanation (3) is consistent with the relatively small maximum particle size ($\beta_\mathrm{min} \sim 0.0005$) and the significantly reduced velocity (width) of large particles ($\beta<0.001$), forming a thin curved western tail.

The terminal velocity must  also depend on the sublimating source area, which sets the length scale over which gas drag can accelerate particles.  Smaller source areas eject particles more slowly than large source areas, all else being equal (Jewitt et al.~2014).  The terminal velocity will be further reduced compared to the ``standard'' picture of a uniform isotropic sublimator when a rotationally modulated gas source is considered. To see this, consider the modulated gas production caused by the diurnal cycle of insolation on a rotating nucleus. Particles lifted from the sublimating day side of the nucleus travel a distance $d \sim V t$ in time $t$. As the Sun sets and the gas flux declines (to zero, at night), these particles will begin to fall back towards the surface under nucleus gravity. Averaged over a nucleus rotation, the net motion will depend on $d$ compared to $r_n$. If $d \ll r_n$, the particle will fall back to the surface every ``day'' and the net speed will be zero, when averaged over the rotation period. If $d \gg r_n$, the particle is launched so high that nucleus gravity cannot pull it back to the surface before the next burst of day-side sublimation occurs, and the particle escapes with a net speed $\sim V$. In the intermediate case, $d \sim r_n$, the particle would rise and fall but there would not be enough time at night for the particle to fall all the way back to the surface. The net speed would then be between 0 and $V$.

For a better understanding of the unique morphology, we created model images of O1 using a Monte Carlo dynamical procedure developed in Ishiguro et al. (2007) and used in Kim et al. (2017).  We assume that dust particles are ejected in a sunward cone with a half-angle $\omega$.  We adopt a decelerated terminal dust velocity ($V_{T}$, as in explanation (3), above) in which larger particles are launched so slowly that they are substantially decelerated by the gravity of the nucleus from which they escape, following:

\begin{equation}
\begin{cases}
V_{ej}  = V_0~a_{\mu m}^{-1/2} \\
V_{T} = \sqrt{V_{ej}^2 - \frac{8\pi G \rho r_*^2}{3}}~\left(V_{ej}^2 > \frac{8\pi G \rho r_*^2}{3}\right)
\end{cases}      
\label{vel}
\end{equation}

\bigskip
\noindent where $V_0$ is the ejection velocity of particles with $a=1~\mu$m. In the absence of an empirical determination, we used the deceleration function in the form of the escape velocity of an object of radius $r_*$ and density $\rho$ = 1000 kg m$^{-3}$.
To reflect uncertain factors such as the location of the dust source on the nucleus, we multiplied the terminal dust velocity $V_T$ by a Gaussian random variable $v$ with the standard deviation $\sigma_v$ = 0.3.
We set the minimum terminal dust speed to zero.
We assumed that the ejected particles follow a differential power-law size distribution with index $q$ = -3.5, minimum particle radius $a_0=0.08$~mm and maximum particle radius $a_1=1.14$~mm. 
The model assumes that dust is ejected continuously from $t_0$ to $t_1$, where $t_0$ is the time elapsed between the start of dust ejection and the observation, and $t_1$ is the time elapsed between the end of dust ejection and the observation.
We assumed a dust production rate $\propto r_H^{-2}$, where $r_H$ is the  heliocentric distance. 

\begin{figure}
\epsscale{0.96}
\plotone{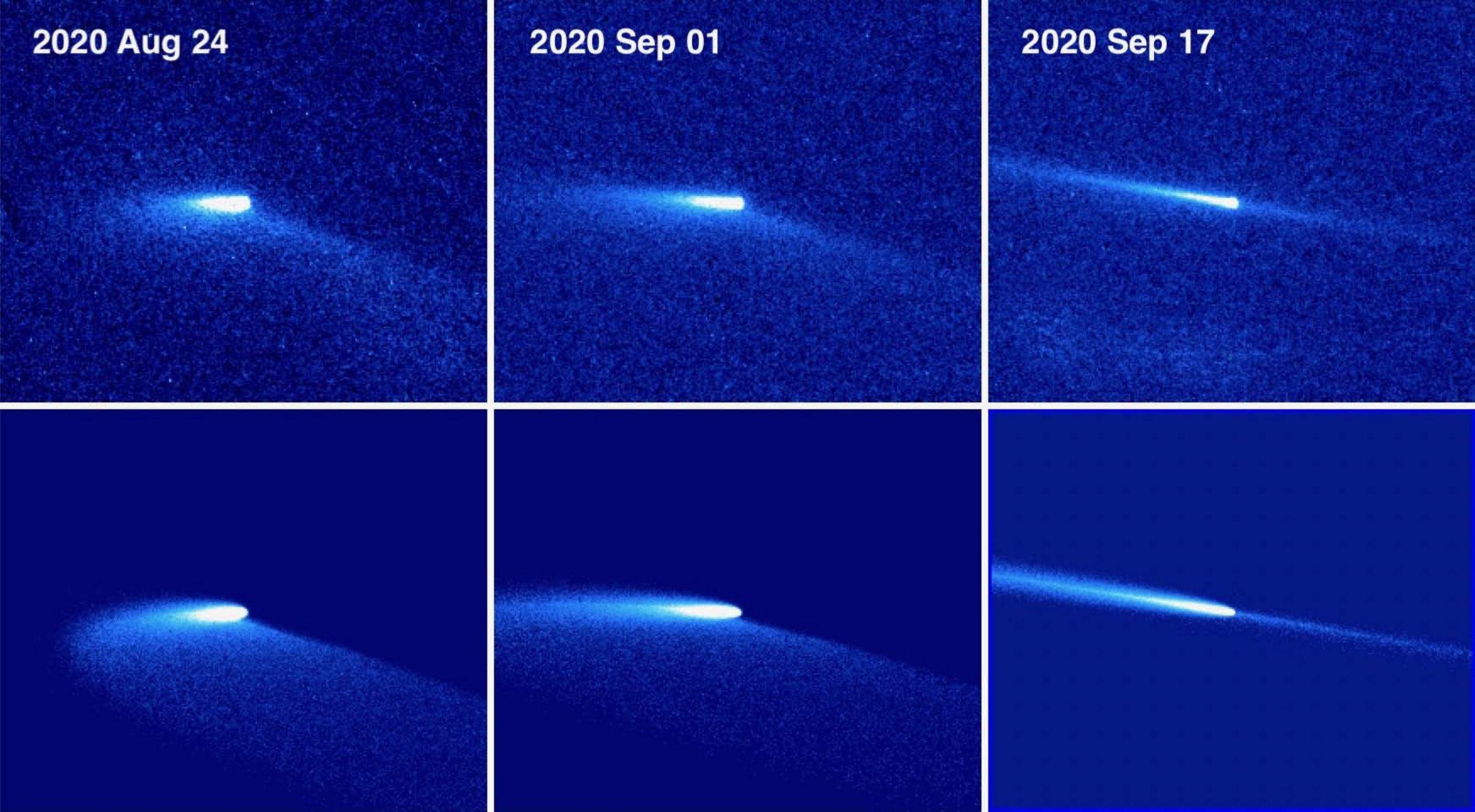}
\caption{Comparison between HST images (upper) and Monte Carlo models (lower) at three epochs of observation. Adopted model parameters are described in Section \ref{model_text}. Celestial north points up and east points to the left.
 \label{model}}
\end{figure}

We created a number of model images using a wide range of parameters as listed in Table \ref{tab:parameter}, and fitted the image from the east parts to the west parts. The model images were visually compared to the observations to find plausible parameters, and then we used least-squares fitting of the eastern tail isophotes to find the best-fit parameters.  
The parameters that fit the eastern tail without deceleration could not simultaneously reproduce the small vertical extent of the western tail. Only when the deceleration parameters (Equation \ref{vel}) were used, could the observed width and morphology  be reproduced.
We conclude that the data are most consistent with explanation (3).
Figure \ref{model} compares the observations with the models on three representative dates.  The best-fit parameters are given in Table \ref{tab:parameter}.

\begin{figure}
\epsscale{1.3}
\plotone{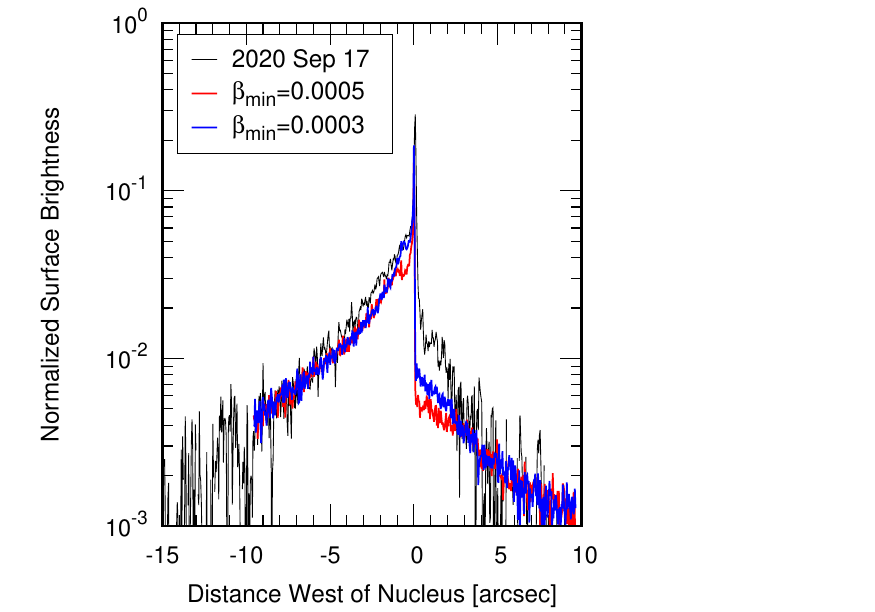}
\caption{Measured surface brightness along the tail (black line), with sample models showing the effect of maximum particle size ($\beta_\mathrm{min}$).
The size distribution index ($q$) does not significantly affect the surface brightness gradient of the west tail.
\label{sbr}}
\end{figure}

Figure \ref{sbr} shows the surface brightness (black line) as a function of angular distance from the nucleus along the tail using data from UT 2020 September 17. The surface brightness was averaged over a region extending $\pm$0.8\arcsec~from the axis, with the sky background determined from the average of adjacent rectangles 1.6\arcsec~above and below the tail.
The model (Figure \ref{sbr}) shows that the slope of the surface brightness profile west of the nucleus is controlled by $\beta_\mathrm{min}$, with acceptable fits requiring $\beta_\mathrm{min} \lesssim$0.0003 (blue line).

On the other hand, the curved west tail morphology require $\beta_\mathrm{min} \gtrsim$0.0005 to fit the data (Figure \ref{synsyns}).
We note that improving the fit without introducing more parameters is beyond our point, and that the actual size distribution will be more complex than the single power-law.
We find that the size distribution index ($q$) only affects the surface brightness gradient of the east tail and not the west tail.

\subsection{Activity Mechanisms}
\label{mechanism}

The presence of a persistent anti-solar tail at the time of each observation indicates that the mass loss of O1 is continuous. As suggested by the synchrone analysis (Figure \ref{synsyns}), the onset must be mid-May 2020 or earlier to form a western tail of detected length. Our best-fit model solution also supports onset in May. As of 2020 September 17, the observed dust must be continuously emitted for at least 130 days to fit the observed data. Onsets earlier than this date tune the shape of the west tail edge, but give only minor differences and do not affect the detectable tail surface brightness profile. Photometry infers that the termination point of the activity is around November-December 2020 (Section \ref{photometry}). The presence of an anti-solar tail in the November 25 data and the absence of a near-nucleus coma gap in the dust tail surface brightness profile suggest that activity continued until the end of November. From 2020 May to 2020 November, the period of inferred activity of O1 is 6 months, or 180 days.

The onset of the mass loss near perihelion, its protracted nature, and its termination at true anomaly $\sim$60$\degr$ are all compatible with a sublimation origin.  
In contrast, the distribution of the dust does not simply follow any of the synchrones (Figure \ref{synsyns}), and so is inconsistent with impulsive ejection as would be expected, for example, for an impact origin.  
Also, the models show that a velocity law of the form  $V \propto a^{-1/2}$  is needed  to fit the flaring east tail of width $w_T \propto \ell_T^{1/2}$  (Figure \ref{fwhm}), as expected of gas drag particle acceleration. Continuous emission with size-independent velocity ($V \propto a^{0}$) cannot reproduce the measured tail width.  Sublimation offers a natural explanation of both the protracted period of the mass loss and the flared dust distribution projected to the east of the nucleus.

We estimate an order of magnitude dust production rate using

\begin{equation}
\frac{dM}{dt} = \frac{4}{3}\frac{\rho \overline{a} C_d}{\tau_r},
\label{dmbdt}
\end{equation}

\noindent where $\rho$ = 1000 kg m$^{-3}$ is the assumed particle density,  $\overline{a}$ is the mean particle radius, $C_d = C_e - C_n$ is the dust cross-section in a photometric aperture (Table 2), and $\tau_r$ is the residence time in the aperture.
The cross-section within the $L$ = 8,000~km radius aperture is $\sim$7 km$^2$ in 2020 August.
The residence time $\tau_r$ is given by $(2L/\beta g_{\odot})^{1/2}$, where we take the solar gravitational acceleration $g_{\odot}$ = 10$^{-3}$ m s$^{-2}$ and $\beta \sim 0.002$ to find $\tau_r \sim$ 3$\times 10^6$~s.
With $\overline{a}$ = 0.3 mm, Equation (\ref{dmbdt}) gives dust production rates $dM/dt \sim$ 0.9 kg s$^{-1}$ in 2020 August declining to nearly zero by 2020 December.

We solved the energy balance equation for an exposed water ice surface located at the sub-solar point on O1. At $r_H$ = 2.37 AU ($T$ = 192 K), we find that ice would sublimate, in equilibrium with sunlight, at the specific rate $F_s$ = 5.7$\times$10$^{-5}$ kg m$^{-2}$ s$^{-1}$.  The area of exposed ice needed to supply a dust production rate, $dM/dt \sim$ 0.9 kg s$^{-1}$, is given by 

\begin{equation}
A_s = \frac{dM/dt}{f_{dg} F_s}
\label{subl_area}
\end{equation}

\noindent where $f_{dg}$ is the ratio of dust to gas mass production rates.  We adopt $f_{dg}$ = 10 (Fulle et al. 2016; Reach et al. 2000) to find $A_s$ = 1580 m$^2$ (only $\sim$0.07\% of the surface of a spherical nucleus of radius 420 m) corresponding to a circular patch as small as $r_s = (A_s/\pi)^{1/2}$ $\sim$22~m in radius.

\begin{figure}
\epsscale{1.3}
\plotone{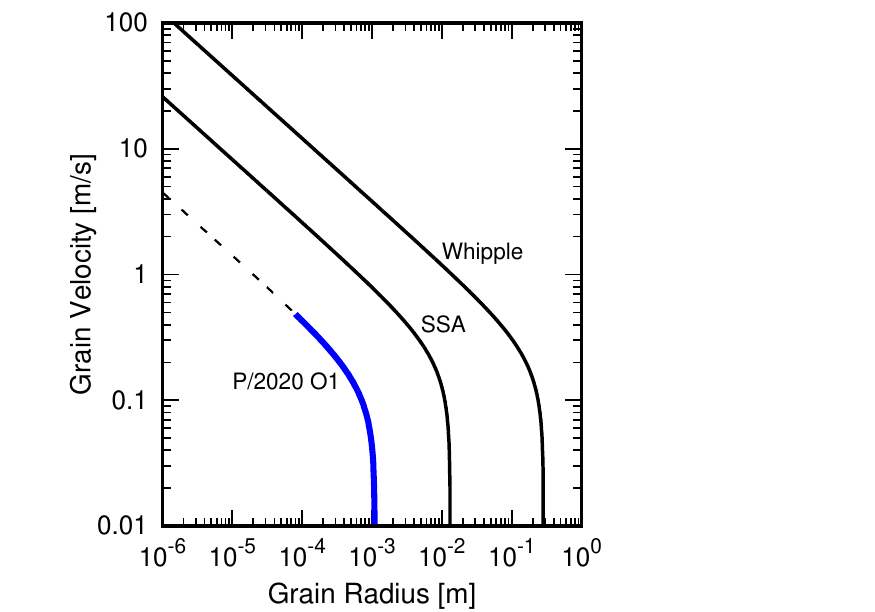}
\caption{The blue curve shows the empirical dust grain ejection velocity from O1 as a function of particle radius.  Black curves show the velocity-radius relation from the Small Source Approximation (SSA) and the Whipple model, respectively, as described in the text.
 \label{whipple}}
\end{figure}

Figure \ref{whipple} shows models of the dust grain ejection velocity from O1 as a function of particle radius.  The blue curve shows the best-fit dust terminal velocity obtained from the Monte Carlo model (Equation \ref{vel}).  For comparison, we show the velocities predicted using the classical comet model (Whipple 1951) and the small source approximation (SSA) model (Jewitt et al.~2014). In the latter case, a small sublimating area (length scale $r_s \sim$ 22 m, as above) limits the acceleration length for gas-entrained dust particles, resulting in lower ejection speeds.

The measured dust speeds are smaller than those of the classical model by about an order of magnitude and  smaller than predicted by the SSA model by a factor $\sim$3 (Figure \ref{whipple}).
Ejection velocities of 0.1 mm particles are comparable to the 0.3 m s$^{-1}$ gravitational escape speed of the (non-rotating) nucleus, while larger particles are launched at speeds less than the gravitational escape velocity.
Similarly low velocities have been reported in other active asteroids (133P, 313P, and 288P) and explained as centripetal-assisted sublimation (Jewitt et al. 2014, 2015; Agarwal et al. 2016).
Like in 133P,  the sublimation of near-surface ice aided by centripetal acceleration may be responsible.
Our tentative evidence for rapid rotation of the nucleus (Figure \ref{lightcurve}) is consistent with this hypothesis.

While sublimation provides the most plausible explanation for the activity of O1, we need additional observations to demonstrate the expected recurrence of activity at subsequent perihelia.
O1 will next reach perihelion in 2024 August, and observations should be made near this time to search for the recurrence of activity that is the hallmark of the MBCs.
The current inner edge of the MBC population appears to be $a \sim$ 2.75 AU at which MBC 259P (Hsieh et al. 2021) and the outgassing dwarf planet (1) Ceres (K{\"u}ppers et al. 2014) are located.
If the activity is repetitive near perihelion and water ice sublimation is thus confirmed, O1 would be the innermost known MBC ($a$ = 2.647 AU), and it would allow us to extend the inner edge of the ice-bearing asteroid zone inward by $\sim$0.1 AU.

\clearpage

\section{SUMMARY}
We present  Hubble Space Telescope measurements of active asteroid P/2020 O1.
We find that

\begin{enumerate}

\item The nucleus of O1 has absolute magnitude $H_V$ = 19.25$\pm$0.13.  With assumed geometric albedo $p_V = 0.05$, the equivalent circular radius is $r_e = 420$ m. 
Time-series photometry shows tentative evidence for the rapidly rotating nucleus (double-peaked period $\sim$1.67 hr).

\item We explored a range of Monte Carlo models to fit the eastern tail isophotes, width, and curved western tail morphology.
The best-fit models involve continuous emission, over a period of 180 days from 2020 May to 2020 November, of dust grains (radius $\sim$0.08--1.14 mm) with terminal ejection velocities of $V_T \sim$ 0--0.3 m s$^{-1}$.

\item The onset of the mass loss near perihelion, its protracted nature,  its termination at true anomaly $\sim$60$\degr$ and the $V \propto a^{-1/2}$ functional dependence of the dust velocity on particle radius are all compatible with a sublimation origin.   

\item The peak mass loss rate in dust is $\sim$0.9 kg s$^{-1}$, decreasing on an e-folding time $\sim$70 days. 
Equilibrium sublimation of exposed water ice from as little as 1580 m$^2$ ($\sim$0.07\% of the nucleus surface) can account for the detected mass loss.

\item Ejection velocities of $\sim$0.1 mm particles are comparable to the 0.3 m s$^{-1}$ gravitational escape speed of the nucleus, while larger particles are released at speeds less than the  escape velocity from a non-rotating body. These properties are consistent with the sublimation of near-surface ice assisted by the centripetal acceleration of the rapidly rotating nucleus.

\end{enumerate}

\acknowledgments
We thank the anonymous referee for a prompt review.
Based on observations made under GO 16308 and 16463 with the NASA/ESA Hubble Space Telescope, obtained at the Space Telescope Science Institute,  operated by the Association of Universities for Research in Astronomy, Inc., under NASA contract NAS 5-26555.    Y.K. and J.A. acknowledge funding by the Volkswagen Foundation. J.A.'s contribution was made in the framework of a project funded by the European Union's Horizon 2020 research
and innovation program under grant agreement No 757390 CAstRA.

{\it Facilities:}  \facility{HST}.



\begin{deluxetable}{lccccrccccr}
\tablecaption{Observing Geometry 
\label{geometry}}
\tablewidth{0pt}
\tablehead{ \colhead{UT Date and Time}   & \colhead{DOY\tablenotemark{a}} & $\Delta T_p$\tablenotemark{b} & $\nu$\tablenotemark{c} & \colhead{$r_H$\tablenotemark{d}}  & \colhead{$\Delta$\tablenotemark{e}} & \colhead{$\alpha$\tablenotemark{f}}   & \colhead{$\theta_{-\odot}$\tablenotemark{g}} &   \colhead{$\theta_{-V}$\tablenotemark{h}}  & \colhead{$\delta_{\oplus}$\tablenotemark{i}}   }
\startdata
2020 Aug 24  10:42 - 11:23 & 237 & 114 & 32.7 & 2.369 & 1.443 & 12.6 & 87.5 & 260.5 & -1.5\\
2020 Sep 01  14:11 - 14:52 & 245 & 122 & 35.0 & 2.375 & 1.506 & 15.5 & 84.5 & 260.7 & -0.9\\
2020 Sep 17  17:50 - 18:31 & 261 & 138 & 39.5 & 2.387 & 1.662 & 20.1 & 81.0 & 260.8 & -0.0\\
2020 Nov 25  06:15 - 06:56 & 330 & 207 & 58.3 & 2.453 & 2.549 & 22.6 & 72.2 & 257.2 & 1.8\\
2020 Dec 25  07:36 - 08:18 & 360 & 237 & 66.1 & 2.487 & 2.927 & 18.7 & 69.3 & 254.8 & 1.7\\
2021 Jul 22  02:19 - 06:08 & 569 & 446 & 114.9 & 2.745 & 3.026 & 19.4 & 254.7 & 259.6 & -1.4
\enddata

\tablenotetext{a}{Day of Year, UT 2020 January 01 = 1}
\tablenotetext{b}{Number of days from perihelion (UT 2020-May-03 = DOY 124).}
\tablenotetext{c}{True anomaly, in degrees}
\tablenotetext{d}{Heliocentric distance, in AU}
\tablenotetext{e}{Geocentric distance, in AU}
\tablenotetext{f}{Phase angle, in degrees}
\tablenotetext{g}{Position angle of the projected anti-Solar direction, in degrees}
\tablenotetext{h}{Position angle of the projected negative heliocentric velocity vector, in degrees}
\tablenotetext{i}{Angle of Earth above the orbital plane, in degrees}

\end{deluxetable}

\clearpage
\begin{deluxetable}{lccccccc}
\tabletypesize{\scriptsize}
\tablecaption{Photometry with Fixed Linear Apertures
\label{phot}}
\tablewidth{0pt}
\tablehead{
\colhead{UT Date}    & \colhead{Quantity\tablenotemark{a}} & \colhead{500 km }   & \colhead{1000 km} & \colhead{2000 km} & \colhead{4000 km} & \colhead{8000 km}
}

\startdata

2020 Aug 24 	&	$V$ 		& 21.45$\pm$0.03	& 21.01$\pm$0.02	 & 20.55$\pm$0.02	 & 20.14$\pm$0.01	& 19.84$\pm$0.01 \\
2020 Aug 24      &	$H$ 		&18.03$\pm$0.03	& 17.60$\pm$0.02	 & 17.13$\pm$0.02	 & 16.72$\pm$0.01	& 16.42$\pm$0.01 \\
2020 Aug 24 	& 	$C_e$ 	& 1.69$\pm$0.05 	& 2.52$\pm$0.05	& 3.88$\pm$0.07	& 5.66$\pm$0.05	& 7.43$\pm$0.07 \\\\

2020 Sep 01   	 &	$V$ 		& 21.87$\pm$0.03	& 21.43$\pm$0.03	& 20.98$\pm$0.02	& 20.59$\pm$0.02	 & 20.29$\pm$0.02 \\
2020 Sep 01 	 &	$H$	 	& 18.24$\pm$0.03	 & 17.81$\pm$0.03	 & 17.36$\pm$0.02	& 16.97$\pm$0.02	& 16.67$\pm$0.02 \\
2020 Sep 01 	& 	$C_e$ 	& 1.39$\pm$0.04	& 2.08$\pm$0.06	 & 3.15$\pm$0.06	& 4.50$\pm$0.08	 & 5.94$\pm$0.11 \\\\

2020 Sep 17 	 &	$V$ 		& 22.28$\pm$0.04	 & 21.86$\pm$0.03	 & 21.38$\pm$0.02	& 20.86$\pm$0.02	& 20.50$\pm$0.02 \\ 
2020 Sep 17 	 &	$H$	 	& 18.28$\pm$0.04	 & 17.87$\pm$0.03	 & 17.38$\pm$0.02	& 16.86$\pm$0.02	& 16.50$\pm$0.02 \\
2020 Sep 17 	& 	$C_e$ 	& 1.34$\pm$0.05  	& 1.97$\pm$0.06 	& 3.08$\pm$0.06 	& 4.96$\pm$0.09	& 6.93$\pm$0.13 \\\\

2020 Nov 25 	&	$V$  		& 24.14$\pm$0.09	 & 23.96$\pm$0.08	& 23.54$\pm$0.07	& 23.17$\pm$0.06	& 22.90$\pm$0.05 \\
2020 Nov 25 	 &	$H$	 	& 19.07$\pm$0.09	 & 18.90$\pm$0.08	 & 18.48$\pm$0.07	& 18.11$\pm$0.06	& 17.84$\pm$0.05 \\
2020 Nov 25	& 	$C_e$ 	& 0.65$\pm$0.06 	& 0.76$\pm$0.06 	& 1.12$\pm$0.07 	& 1.57$\pm$0.09 	& 2.02$\pm$0.10 \\\\

2020 Dec 25 	&	$V$  		& 24.32$\pm$0.10	 & 24.37$\pm$0.10	 & 24.20$\pm$0.09	& 24.34$\pm$0.10	 & 24.35$\pm$0.11 \\	
2020 Dec 25 	&	$H$	 	& 19.04$\pm$0.10	 & 19.10$\pm$0.10	& 18.93$\pm$0.09	& 19.07$\pm$0.10	& 19.08$\pm$0.11 \\
2020 Dec 25 	& 	$C_e$ 	& 0.66$\pm$0.06  	& 0.63$\pm$0.06 	& 0.74$\pm$0.06 	& 0.65$\pm$0.06 	& 0.64$\pm$0.07 \\\\

2021 Jul 22 	&	$V$  		& 24.83$\pm$0.13	 & 24.85$\pm$0.12	 & 25.02$\pm$0.13	& 24.93$\pm$0.13	& 25.04$\pm$0.16 \\
2021 Jul 22 	 &	$H$	 	& 19.25$\pm$0.13	& 19.27$\pm$0.12	 & 19.44$\pm$0.13	& 19.35$\pm$0.13	& 19.47$\pm$0.16 \\
2021 Jul 22 	& 	$C_e$ 	& 0.55$\pm$0.07 	& 0.54$\pm$0.06 	& 0.46$\pm$0.06 	& 0.50$\pm$0.06	 & 0.45$\pm$0.07 \\\\
\enddata

\tablenotetext{a}{V = apparent V magnitude, $H$ = absolute magnitude computed assuming a C-type phase function, $C_e$ = effective scattering cross-section in km$^2$.}

\end{deluxetable}

\clearpage 
\begin{deluxetable}{lcccccc}
\tablecaption{Nucleus Photometry
\label{nucleus_phot}}
\tablewidth{0pt}
\tablehead{
\colhead{UT Date}    & \colhead{Mid-point Time\tablenotemark{a}} & \colhead{$V$\tablenotemark{b}}   & \colhead{$H$\tablenotemark{c}} 
& \colhead{$C_e$\tablenotemark{d}} 
}

\startdata

2020 Nov 25 & 06:17	&	24.15$\pm$0.06	&	19.09$\pm$0.06	&	0.64$\pm$0.04 	\\
 & 	06:24	&	24.17$\pm$0.07	&	19.10$\pm$0.07	&	0.63$\pm$0.04 	\\
 & 	06:31	&	24.12$\pm$0.06	&	19.06$\pm$0.06	&	0.66$\pm$0.04 	\\
 & 	06:39	&	24.18$\pm$0.07	&	19.11$\pm$0.07	&	0.62$\pm$0.04 	\\
 & 	06:47	&	24.26$\pm$0.08	&	19.19$\pm$0.08	&	0.58$\pm$0.04 	\\
 & 	06:54	&	24.33$\pm$0.08	&	19.27$\pm$0.08	&	0.54$\pm$0.04 	\\

2020 Dec 25 & 07:39	&	24.41$\pm$0.09	&	19.14$\pm$0.09	&	0.61$\pm$0.05 	\\
 & 	07:46	&	24.33$\pm$0.08	&	19.06$\pm$0.08	&	0.66$\pm$0.05 	\\
 & 	07:53	&	24.27$\pm$0.07	&	19.00$\pm$0.07	&	0.70$\pm$0.05 	\\
 & 	08:01	&	24.40$\pm$0.09	&	19.13$\pm$0.09	&	0.61$\pm$0.05 	\\
 & 	08:08	&	24.42$\pm$0.09	&	19.15$\pm$0.09	&	0.60$\pm$0.05 	\\
 & 	08:15	&	24.54$\pm$0.10	&	19.27$\pm$0.10	&	0.54$\pm$0.05 \\

\enddata

\tablenotetext{a}{Exposure Mid-point Time.  }
\tablenotetext{b}{Apparent V magnitude within a 0.2\arcsec~radius aperture.  
} 
\tablenotetext{c}{Absolute magnitude computed assuming a C-type phase function.}
\tablenotetext{d}{Effective scattering cross-section in km$^2$.}

\end{deluxetable}
\clearpage

\begin{table}
  \caption{Dust Model Parameters}
  \begin{center}
    \begin{tabular}{lllll}
\hline
Parameter   & Parameter Range Explored & Best-fit Values & Unit\\
\hline
$\beta_\mathrm{max}$ & 0.005 to 0.01 & 0.007 & --\\
$\beta_\mathrm{min}$ & 0.0001 to 0.001 & 0.0005 & --\\
$t_0\tablenotemark{a}$ & 100 to 200 with 10 interval & $>110$ & days  \\
$t_1$ & 0 & Fixed & days  \\
$V_0$ & 1.0 to 5.0 with 0.5 interval & 4$\pm$1 & m s$^{-1}$\\
$q$ & 3.5 & Fixed & --\\
$\omega$ & 10 to 60 with 5 interval & 25$\pm$5 & degree\\
$r_*$ & 200 to 450 with 10 interval & 250$\pm$50 & m \\
\hline
    \end{tabular}
  \end{center}
 \label{tab:parameter}
 \tablenotetext{a}{Time elapsed between the start of dust ejection and the observation. We find $t_0$ = 110 days on UT 2020 August 24, increasing with time on later dates.}
\end{table}
\clearpage

\clearpage

\clearpage

\clearpage

\clearpage

\clearpage

\clearpage

\clearpage

\end{document}